\RequirePackage{xspace}

\documentclass[preprint,12pt]{elsarticle}

\usepackage{fullpage}
\usepackage{amssymb}

\usepackage{graphicx}
\usepackage{epstopdf}
\usepackage{bm}
\usepackage{amsmath}
\usepackage{multirow}
\usepackage{amsbsy}
\usepackage{url}

\usepackage{mathptmx}

\begin{document}

\begin{frontmatter}

\title{\boldmath }
\title{\large  \bf\boldmath Comparison of the CPU and memory performance of StatPatternRecognition (SPR) and Toolkit for MultiVariate Analysis (TMVA) }

\author{G.~Palombo}
\address{ California Institute of Technology}
\ead{palombo@cacr.caltech.edu}

\begin{abstract}

High Energy Physics data sets are often characterized by a huge number of events. Therefore, it is extremely important to use statistical packages able to efficiently analyze these unprecedented amounts of data. We compare the performance of the statistical packages StatPatternRecognition (SPR) and Toolkit for MultiVariate Analysis (TMVA). We focus on how CPU time and memory usage of the learning process scale versus data set size. As classifiers, we consider  Random Forests, Boosted Decision Trees  and Neural Networks. For our tests, we employ a data set widely used in the machine learning community,  ``Threenorm'' data set, as well as data tailored for testing various edge cases.  For each data set, we constantly increase its size and check CPU time and memory  needed to build the classifiers implemented in SPR and TMVA.
We show that SPR is often significantly faster and consumes significantly less memory. For example, the SPR implementation of Random Forest is by an order of magnitude faster and consumes an order of magnitude less memory than TMVA on Threenorm data.

\end{abstract}





\end{frontmatter}


\section{Introduction}
\label{intro}

In modern High Energy Physics (HEP) analyses, the use of machine learning techniques has become increasingly common. Machine learning classifiers are  used to separate signal events from unwanted background \cite{Roy}. 
Several HEP experiments are characterized by an extremely large number of events. Therefore, it is crucial to use statistical packages able to efficiently analyze data sets described by million events or even larger. 
Statistical packages widely used among statisticians, such as R or Weka, often implement  many statistical techniques that would be extremely useful in HEP analyses. However, they are hardly efficient in dealing with huge data sets \cite{LHC}.

The two statistical software most used  in the HEP community are StatPatternRecognition (SPR) \cite{SPR} and Toolkit for MultiVariate Analysis (TMVA) \cite{TMVA}. Both packages have been developed within the HEP community and are targeted to HEP statistical analyses.
In this work, we compare  how CPU time and memory usage performance scale to data set size in SPR and TMVA.

A detailed description of the machine learning techniques described in this work can be found in Ref. \cite{Stat}.

\section{SPR and TMVA}
\label{sec:1}

SPR is an open source standalone C++ package that can be run within ROOT \cite{ROOT} or from the command line.  It implements linear and quadratic discriminant 
analysis, logistic regression, binary decision splits, bump hunter, two flavors of decision trees, a feedforward backpropagation
neural net with a logistic activation function, several flavors of boosting
including the arc-x4 algorithm, bagging  and random forest. The package also includes two multiclass methods that allow to use any binary classifier for a multiclass classification problem.
Moreover, it implements algorithms to boost or bag any sequence of classifiers as well as combine classifiers trained on subsets of input variables.

TMVA is an open source project integrated with ROOT. It implements the following multivariate techniques: rectangular cuts, projective and multidimensional likelihood estimators, k-Nearest Neighbor, Fisher and H-matrix discriminants, linear and function discriminant analysis,  multilayer perceptron neural networks, support vector machine, boosted decision trees, random forest, and rulefit. It also allows to boost any classifier.

Both packages implement similar techniques to pre-process the data, such as normalization, principal component analysis, correlations, and cuts. Variable importance in TMVA is estimated for neural networks by calculating the weight of neural network links and for decision trees by calculating the improvement in the classifier performance given by the splits on each variable. In addition to these techniques, SPR also presents other variable importance algorithms that work with any classifier: random permutation of the class label, ``Add N Remove R'', and interactions.  

Finally, SPR  also includes cross-validation techniques, allows to choose among 10 Figures of Merit to optimize the classifier and test its performance, and implements Friedman's machine learning-based Goodness of Fit test \cite{FriedmanNew}.  
     
For a more comprehensive description of SPR and TMVA, please see readme and user's guides included in package distributions\cite{SPR,TMVA}. A useful description to get started with both packages can also be found in Ref. \cite{Tom2}.

Among the HEP community, TMVA is considered  more user friendly and SPR is considered more powerful and faster \cite{LHC, Tom2}.

Although a significant number of classifiers are included in both packages, their implementation can differ, leading to different results \cite{LHC}. Different implementation  implies that, for instance, the best parameters for SPR Boosted Decision Trees (BDT) are not necessarily the best parameters for TMVA BDT \cite{Tom2}. 

 A detailed description of a statistical analysis to observe the single-top signal, using both TMVA and SPR, can be found in Ref. \cite{Tom}. Firstly, the author separately looks for the classifier with the best predictive power from TMVA and from SPR.  BDT are the best classifier for TMVA and Random Forest (RF) for SPR.  Finally, it is shown that SPR RF achieves the best overall predictive power. 

 In Ref. \cite{Tom}, TMVA RF predictive power is the lowest. One of the reasons of that poor behavior could be that TMVA default parameter for RF tree depth equals to 3 (although TMVA user's guide reports that tree depth default value is $10^{5}$). Also, another issue with TMVA RF is that the algorithm automatically sets tree depth value back to 3 when that option is manually set to an extremely large value (such as, for instance, $10^{10}$).  RF typically  needs extremely large trees to perform well. Therefore, it is advisable to specifically check TMVA RF tree structure after the classifier has been built. 

\section{Classifiers and Data Sets}
\label{sec:2}
 
For our tests, we use release 08.02.00  of SPR and  4.0.4 of TMVA. CPU time reported in this work refers to elapsed real time. To estimate memory usage, we consider  Resident Set Size (RSS) \cite{Linux}. Memory usage traces are collected by running the top utility at intervals proportional to the expected length of the process. Maximum value for RSS is reported here. Both packages quickly reach a value close to the maximum and then their memory usage becomes almost constant until the end of the process.  The executables are run on a dedicated machine with no other major task running simultaneously. The machine runs CentOS Linux 5.4 and has 8 Intel(R) Xeon(R) @ 2.33 GHz processors with 8 GB of RAM.

 As classifiers, the attention is restricted to RF, BDT, and Neural Networks (NN).

With regards to the most important parameters, for RF we choose 50 trees, at least 5 events per leaf, and each split chosen among $D/2$ variables randomly selected at each node, where $D$ is the data set dimensionality. 
For BDT, we choose 100 cycles and 10\% of the events as minimum number of events per leaf. 
For NN, layer structure is $D:D:D/2:1$. As suggested by TMVA user's guide \cite{TMVA}, TMVA neural network model is the MultiLayer Perceptron (MLP). All other parameters  are set so that the packages would perform the same task for each classifier.   

Goal of the analysis is to check how CPU time and memory usage scale to data set size. The chosen parameter values are not necessarily the best ones for each data set in terms of classifier predictive power. Finding the best classifier configuration for each data set would make hard to exactly keep trace of CPU time and memory usage with respect to the data set size. However, the chosen parameters are likely to be reasonably close to the best possible configuration for every classifier \cite{Talk,Breiman}.

TMVA's variable transformation options are disabled in order to achieve a major similarity between the tasks executed by the packages. 
 TMVA's variable transformation options, which are on by default in the code example, include decorrelation, principal component decomposition, and gaussianisation. They consume a very large amount of memory. For instance, those variable transformation options take up to 5 or 6 times the maximum amount of memory needed to build  BDT. The bigger is the data set, the larger is the  ratio between the memory needed for variable transformation and for training the BDT classifiers.       
SPR does not need such an adjustment since uses different executables to perform different tasks.

As a benchmark data set, we choose  the ``Threenorm'' data set introduced by Breiman in Ref.  \cite{Breiman}. Given a$=2/(20)^{1/2}$, one class is drawn from a unit multivariate normal with mean (a,-a, a, -a, ....,a). The other class is drawn with equal probability from a unit multivariate normal with mean (a,a,...,a) and from a unit multivariate normal with mean (-a,-a,...,-a). 
This data set is considered a difficult one for classification problems \cite{Breiman} and has been widely used for comparison of machine learning algorithms \cite{Comp}.

Moreover, we train our classifiers on edge cases. That is, data sets described by a single variable which follows an unusual distribution. Example of edge cases are a random noise variable, a variable for which all events except for one have the same value, and a variable for which 50\% of events are useless and 50\% have discriminant power. With respect to edge cases, we only test the behavior of RF and BDT.

 Results refer to the training part only. We start with small data sets and gradually increase dimensionality and number of events. We start with $2 \times 10^4$ events and 20 variables.  For BDT, we gradually increase dimensionality up to 100 variables and size up to $4\times 10^6$ events. For RF, being slower given the optimization parameters, we increase up to 100 variables and $10^6$ events; and for NN, the slowest among the three, up to 70 variables and $10^6 $ events. For edge cases, we only increase the number of events as described above.

\begin{figure}[t]
\begin{minipage} [b]{0.5\linewidth}\centering
\includegraphics[scale=0.34, angle=270]{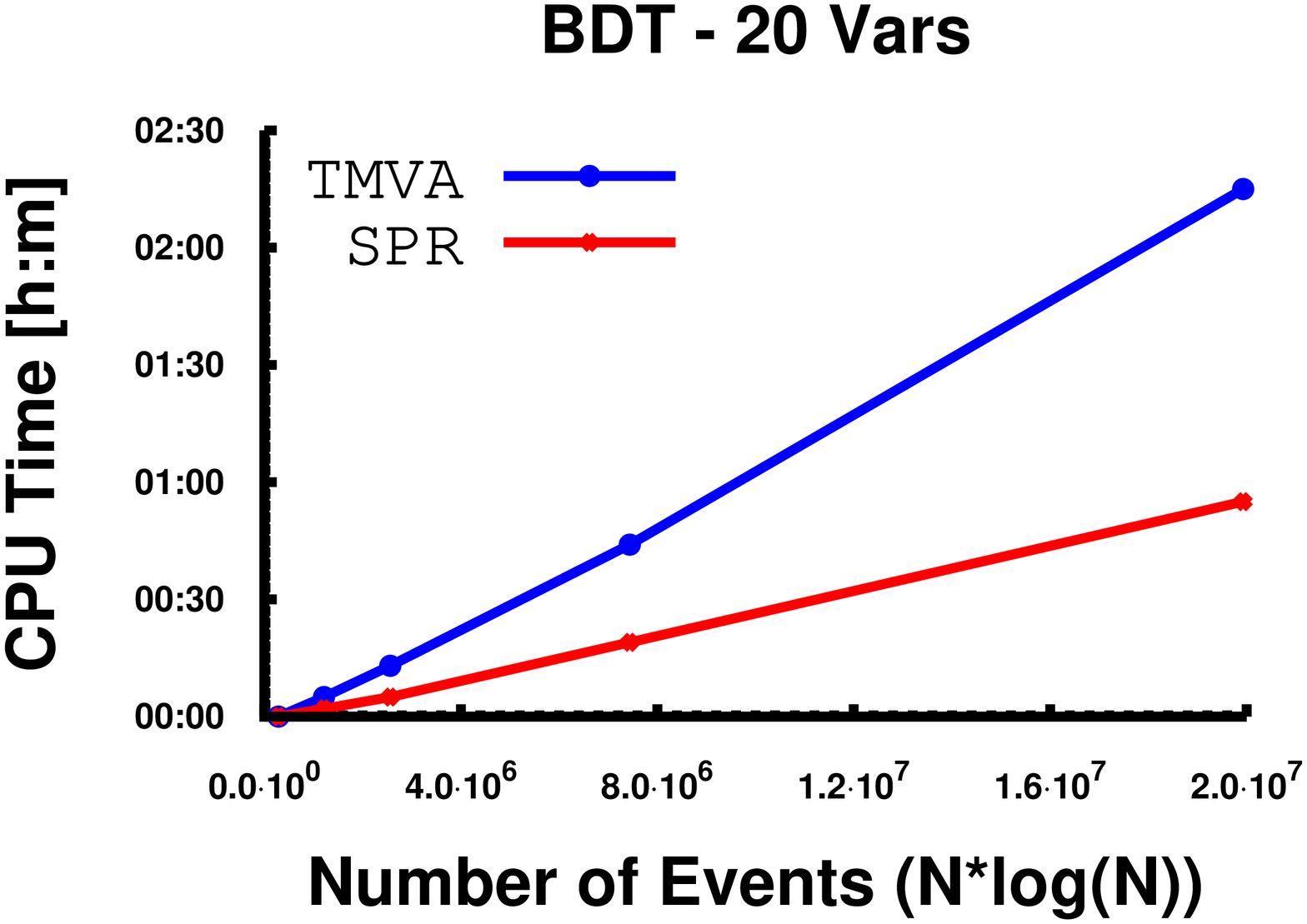}
\end{minipage}
\begin{minipage}[b] {0.5\linewidth} \centering
\includegraphics[scale=0.34, angle=270]{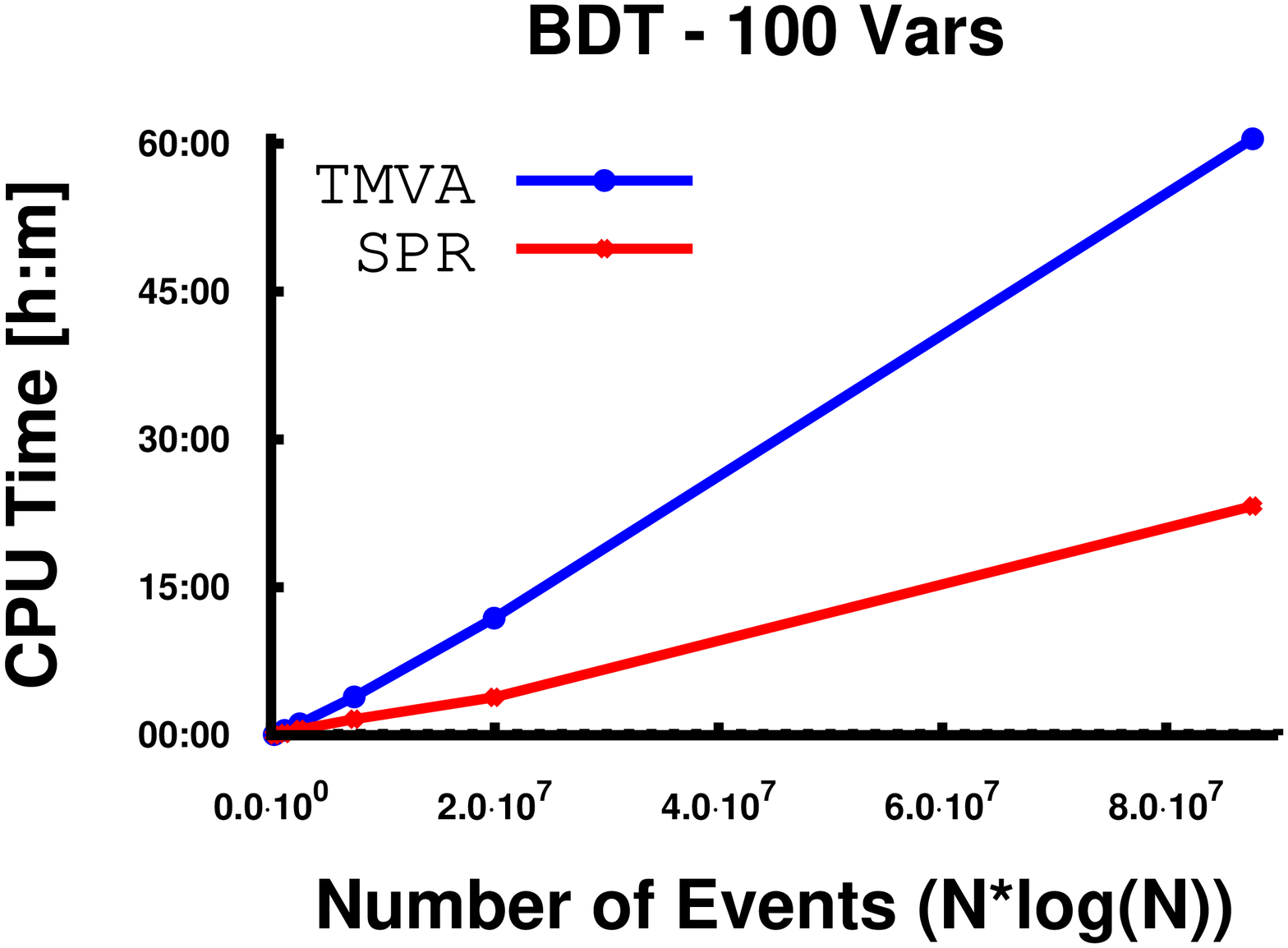}
\end{minipage}
\caption{Threenorm data set: BDT learning CPU time vs number of events for 20 variables (left) and 100 variables (right). }
\label{fig:NormBDTTime}
\end{figure}

\begin{figure}[t]

\begin{minipage} [b]{0.5\linewidth}\centering
\includegraphics[scale=0.34, angle=270]{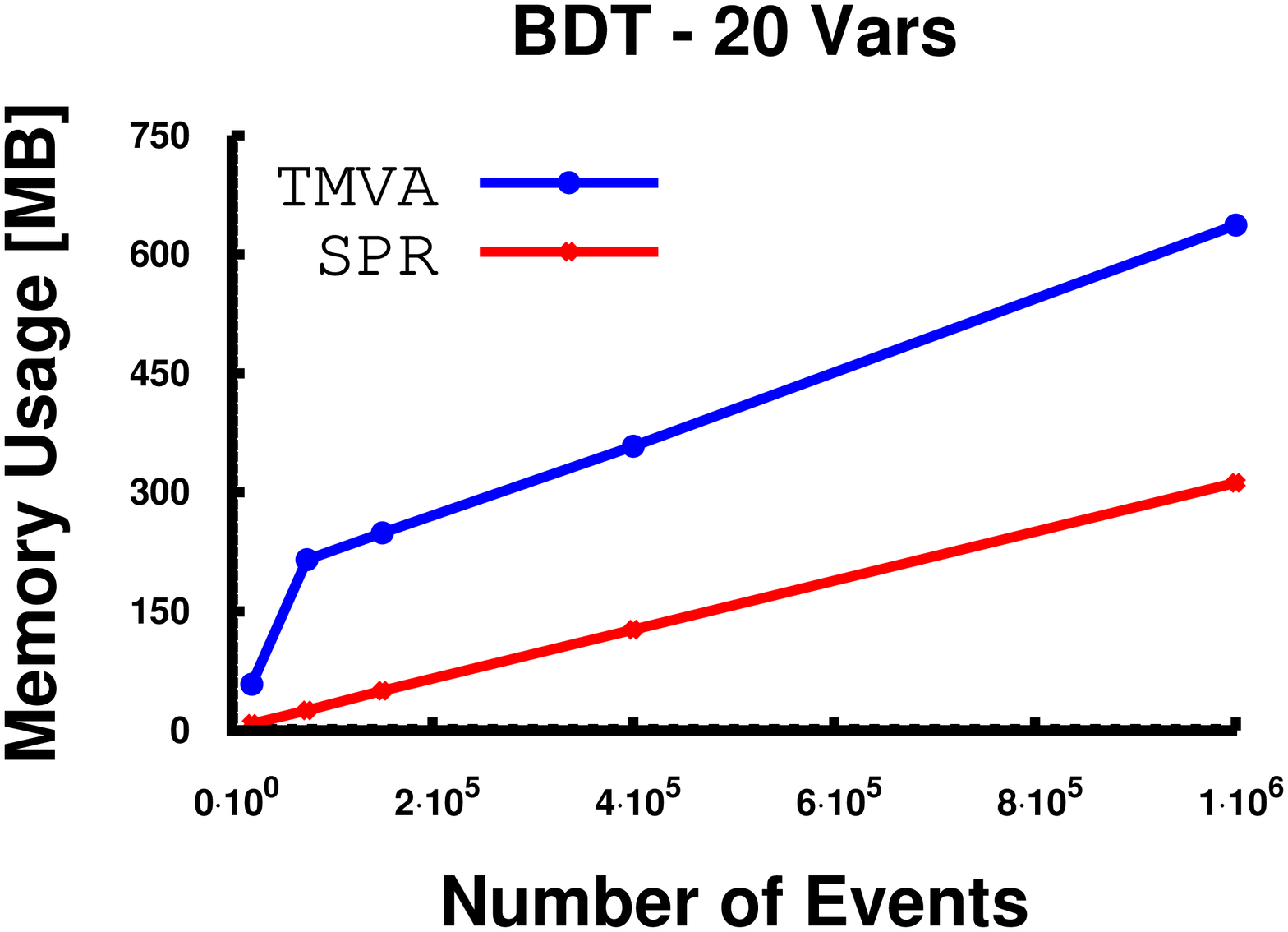}
\end{minipage}
\begin{minipage}[b] {0.5\linewidth} \centering
\includegraphics[scale=0.34, angle=270]{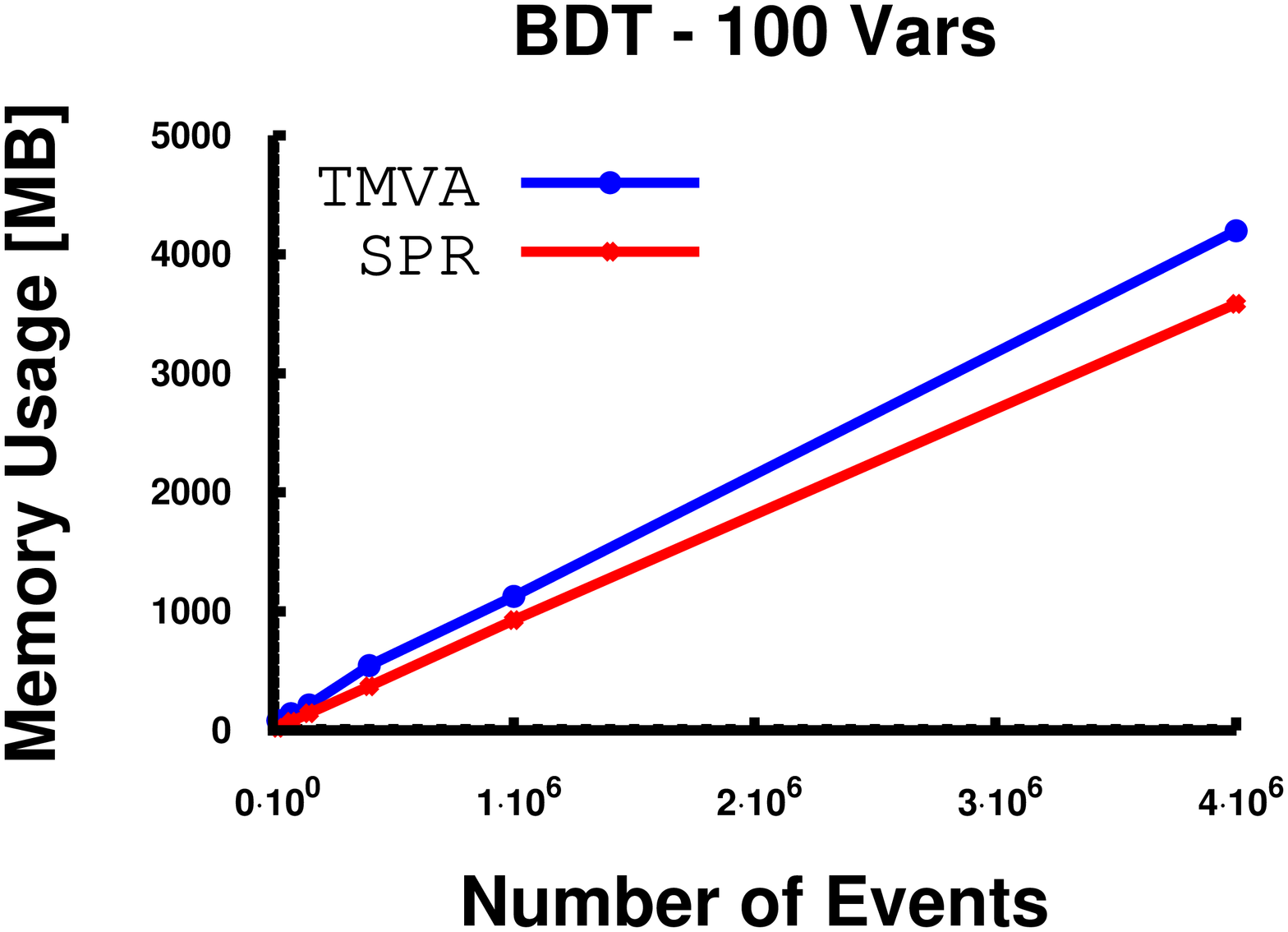}
\end{minipage}
\caption{Threenorm data set: BDT memory usage vs number of events for 20 variables (left) and 100 variables (right).}
\label{fig:NormBDTMemory}
\end{figure}

\section{Test on ``Threenorm'' Data Set}
\label{sec:3}

We test learning time and memory usage on Threenorm data set for BDT, RF, and NN. 
 Fig.~\ref{fig:NormBDTTime} shows a comparison of CPU time needed by SPR and TMVA to build BDT with 20 variables (left) and 100 variables (right). 
 CPU time of tree-based classifiers is expected to grow linearly versus $NlogN$, where $N$ is the number of events. SPR BDT are constantly faster than TMVA BDT. 
We also check configurations with 40 and 70 variables. Results are as expected. That is, they are in between the 20 and 100 variables outputs.

Fig.~\ref{fig:NormBDTMemory} shows SPR and TMVA memory usage for 20 variables and 100 variables. TMVA consumes more memory in both configurations.

The trees built by SPR and TMVA have a very similar structure, with their depth being equal to either 2 or 3. This depends on the optimization parameters chosen, i.e. 10\% of events as  minimum number of events per leaf. Therefore, this analysis gives a useful indication of time and memory difference to build classifiers based on small trees.

\begin{figure}[t]
\begin{minipage} [b]{0.5\linewidth}\centering
\includegraphics[scale=0.34, angle=270]{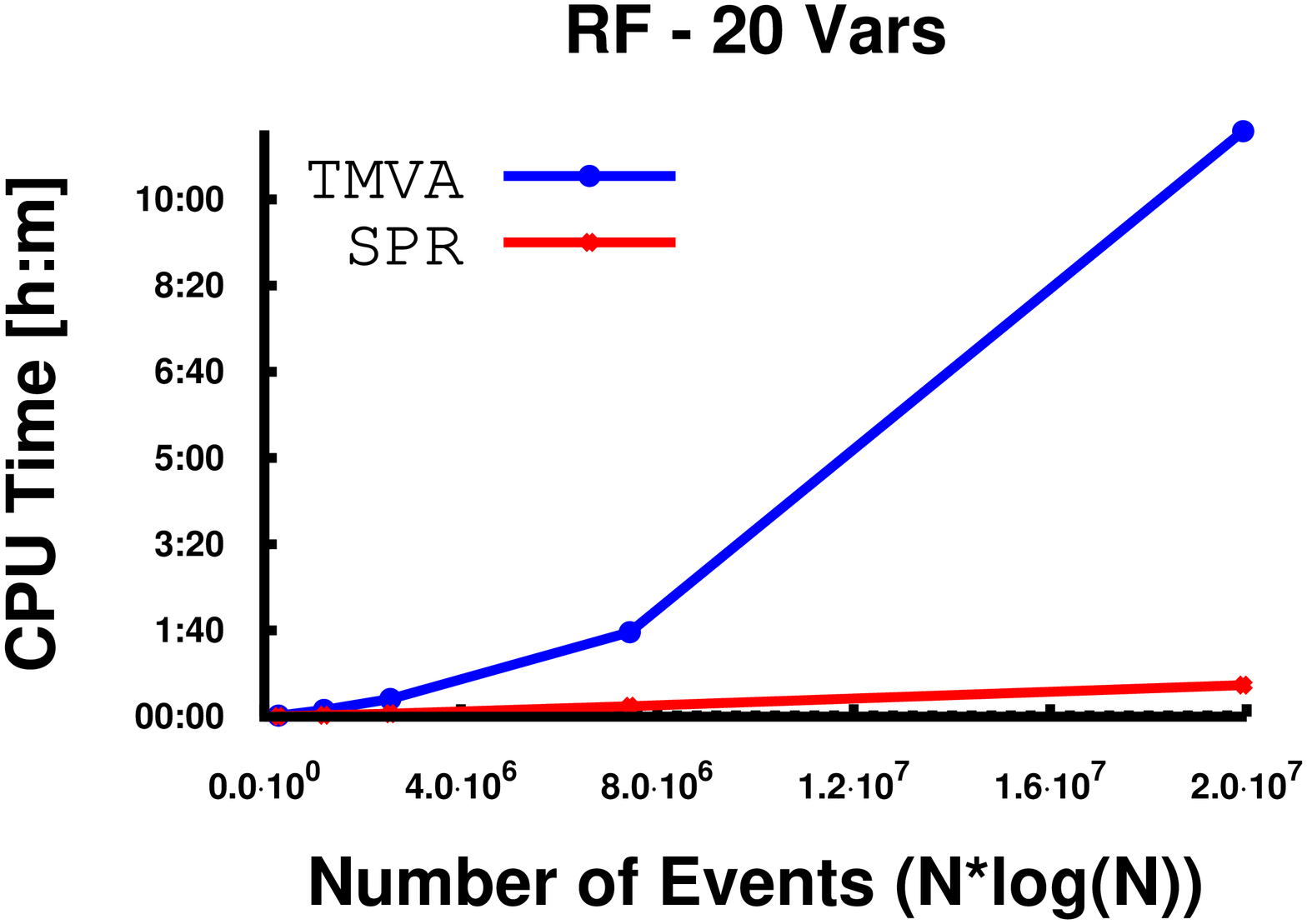}
\end{minipage}
\begin{minipage}[b] {0.5\linewidth} \centering
\includegraphics[scale=0.34, angle=270]{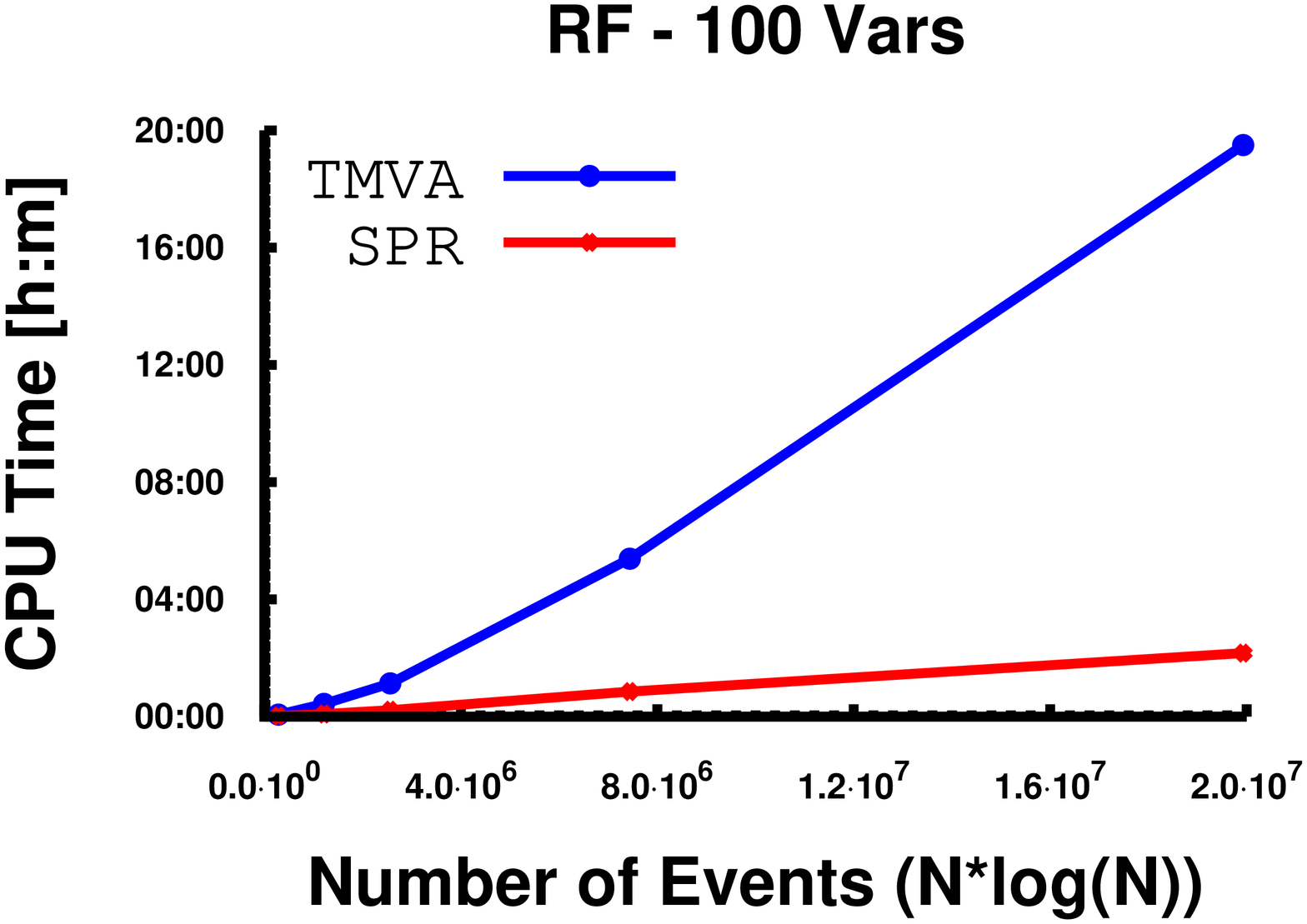}
\end{minipage}
\caption{Threenorm data set: RF learning CPU time vs number of events for 20 variables (left) and 100 variables (right).}
\label{fig:NormRFCPU}
\end{figure}

\begin{figure}[t]
\begin{minipage} [b]{0.5\linewidth}\centering
\includegraphics[scale=0.34, angle=270]{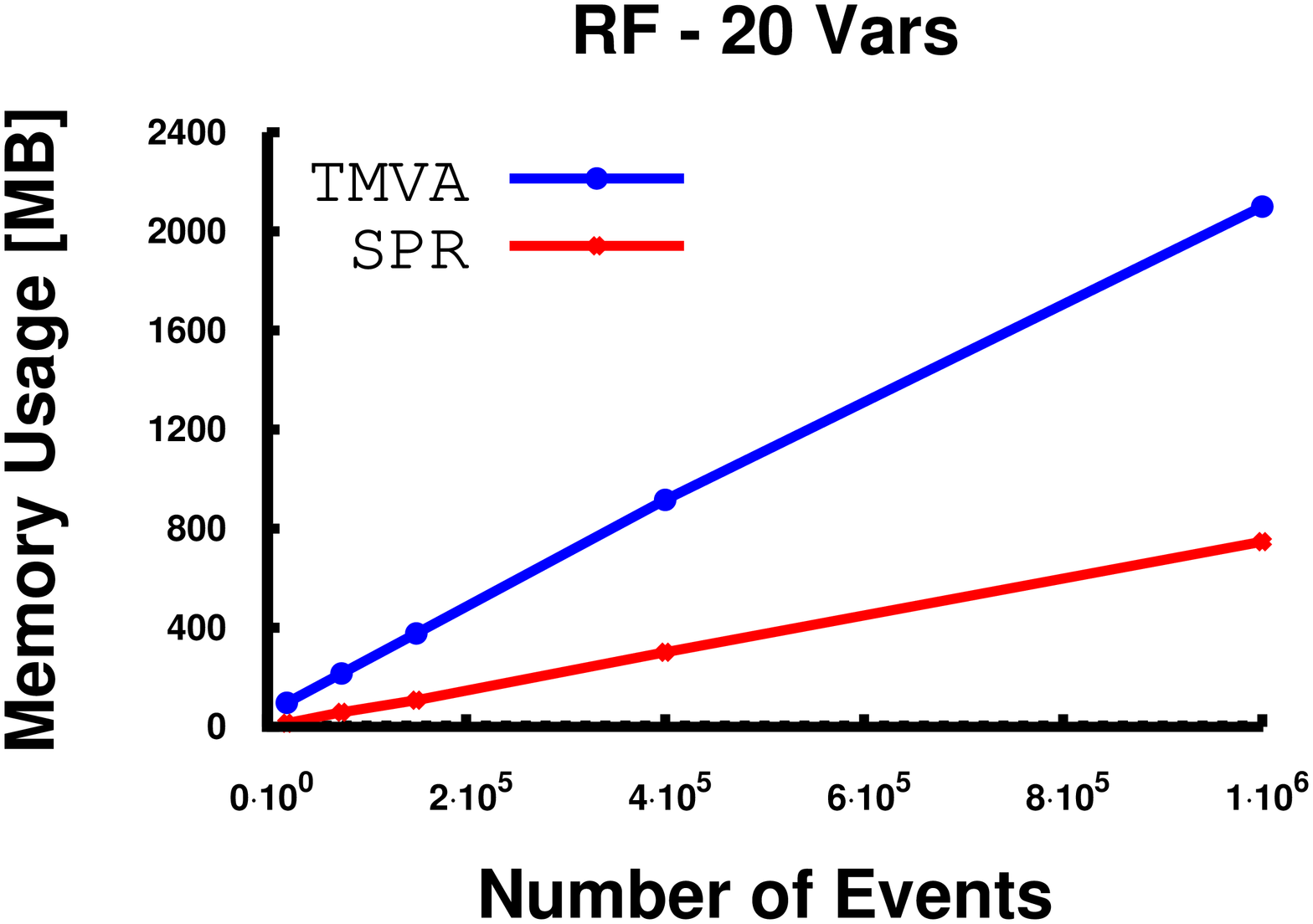}
\end{minipage}
\begin{minipage}[b] {0.5\linewidth} \centering
\includegraphics[scale=0.34, angle=270]{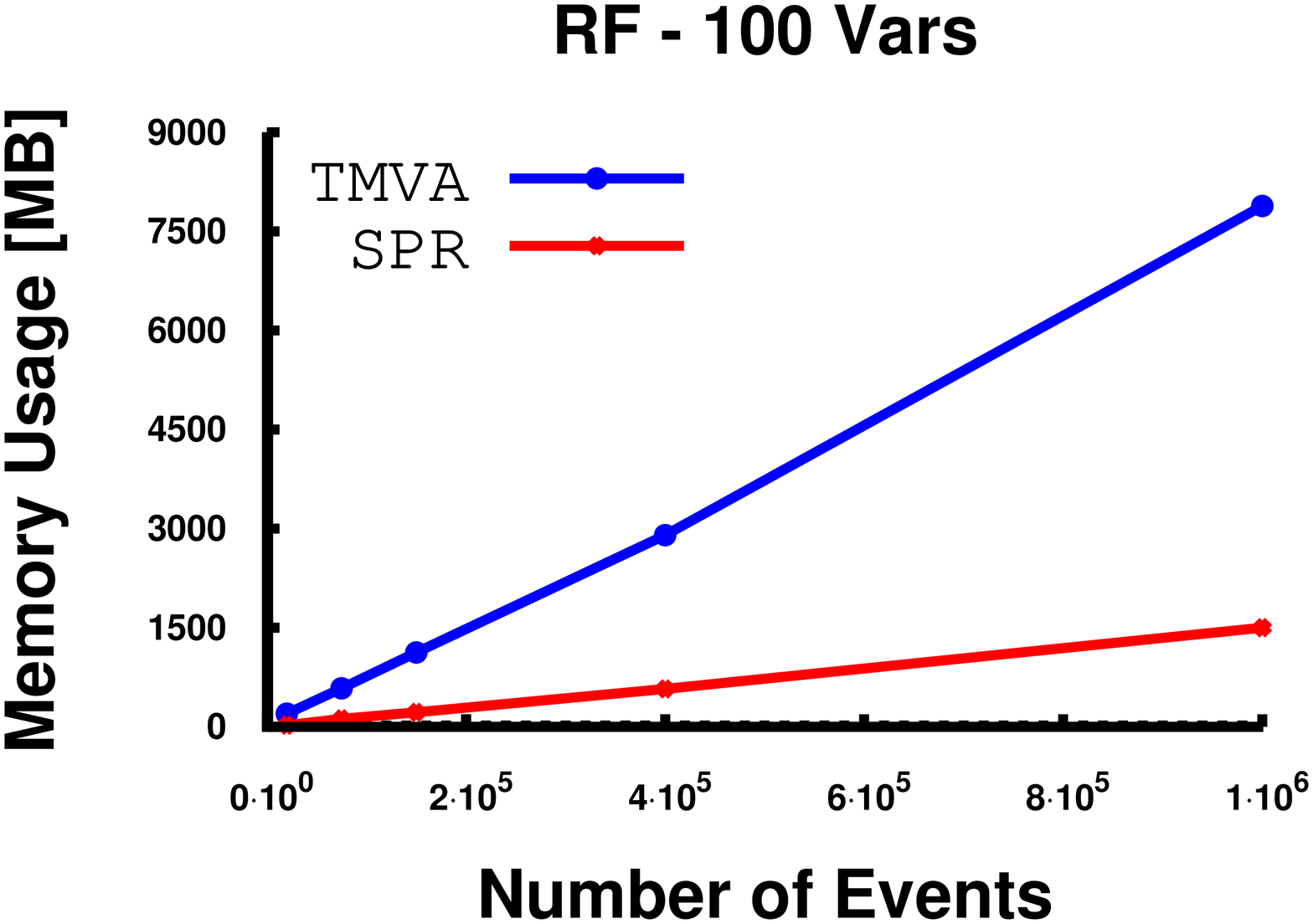}
\end{minipage}
\caption{Threenorm data set: RF memory usage vs number of events for 20 variables (left) and 100 variables (right).}
\label{fig:NormRFMEMORY}
\end{figure}

\newpage

We repeat the same test using RF classifiers. CPU time results are shown in Fig.~\ref{fig:NormRFCPU}. SPR is again faster than TMVA. The difference between SPR and TMVA is larger than for the BDT test.  RF trees are larger than BDT trees. Therefore, bigger trees lead to a larger difference in terms of performance between SPR and TMVA.  
Furthermore, the ratio between TMVA and SPR CPU time increases as the data set size gets larger. That is, SPR CPU time increases linearly with respect to $NlogN$ and TMVA CPU time increases more than linearly.

Fig.~\ref{fig:NormRFMEMORY} shows memory usage results for the RF test. TMVA memory usage is always larger than SPR's one.  

\begin{figure}[t]
\begin{minipage} [b]{0.5\linewidth}\centering
\includegraphics[scale=0.34, angle=270]{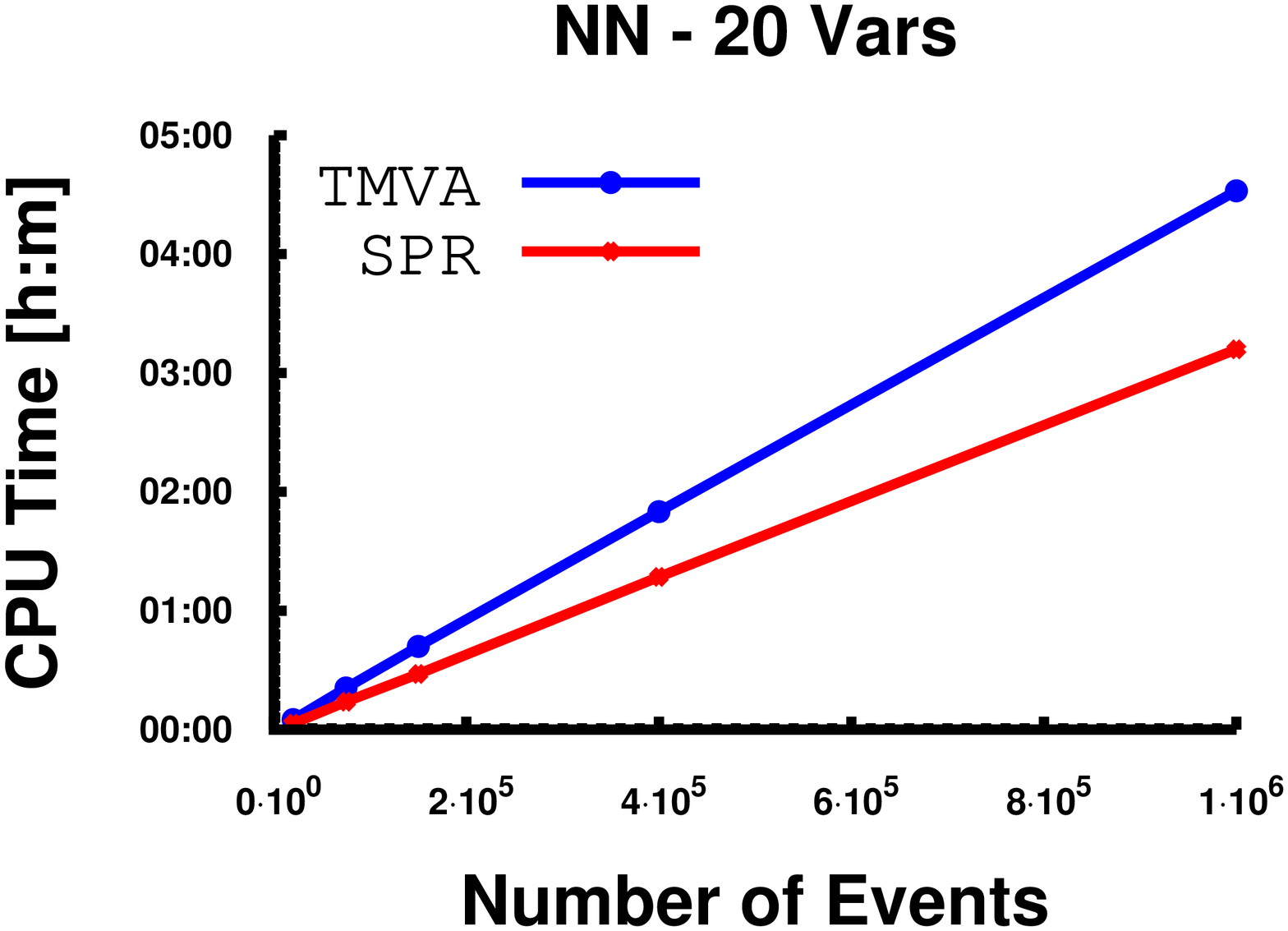}
\end{minipage}
\begin{minipage}[b] {0.5\linewidth} \centering
\includegraphics[scale=0.34, angle=270]{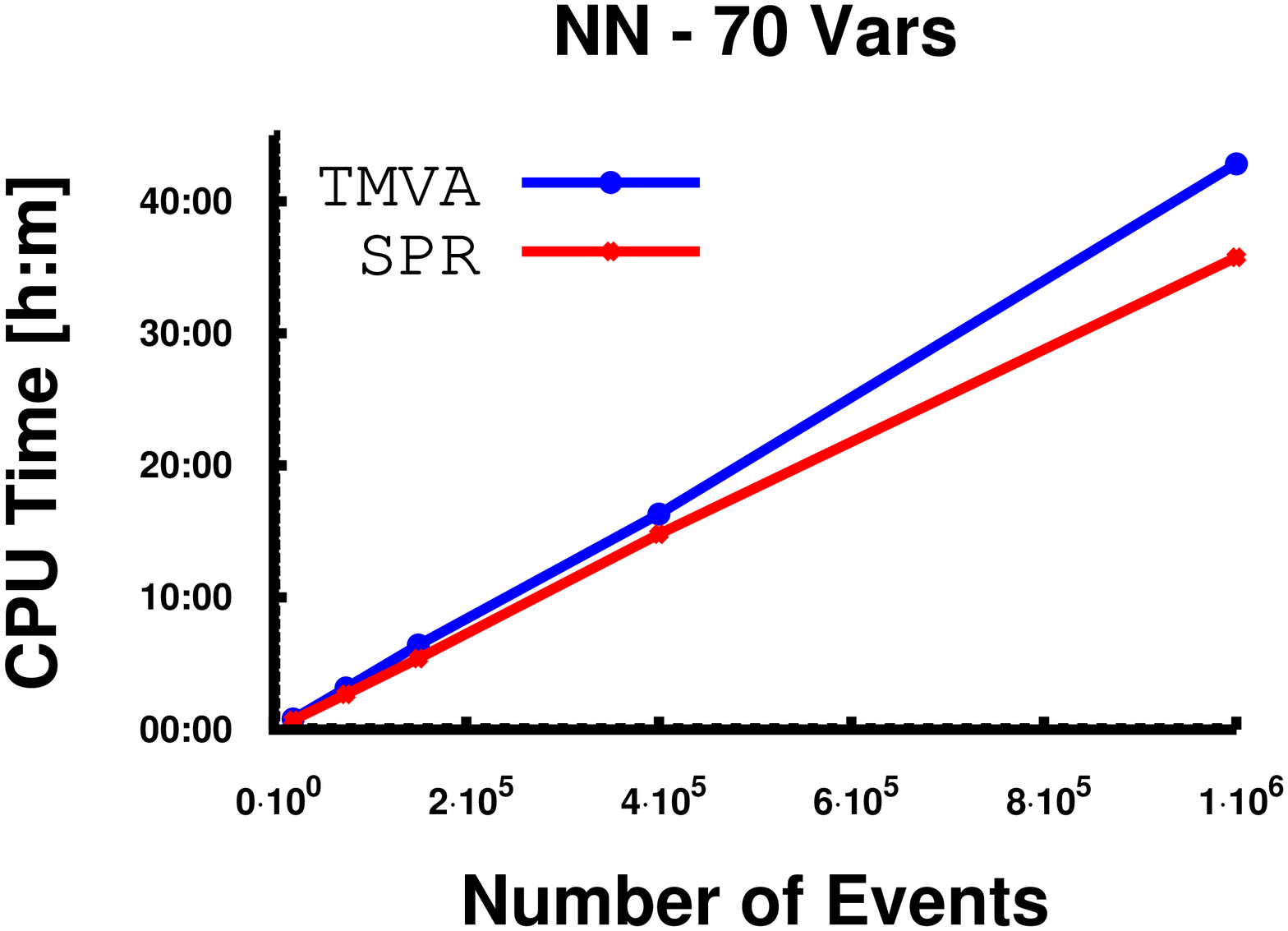}
\end{minipage}
\caption{Threenorm data set: NN learning CPU time vs number of events for 20 variables (left) and 70 variables (right).}
\label{fig:NormNNCPU}
\end{figure}

SPR and TMVA NN CPU time is shown in Fig.~\ref{fig:NormNNCPU} for 20 variables (left) and 70 variables (right). Again, SPR is always faster than TMVA for any data set size. However, the difference between SPR and TMVA is not as large as in the BDT and RF cases.

\begin{figure}[t]
\begin{minipage} [b]{0.5\linewidth}\centering
\includegraphics[scale=0.34, angle=270]{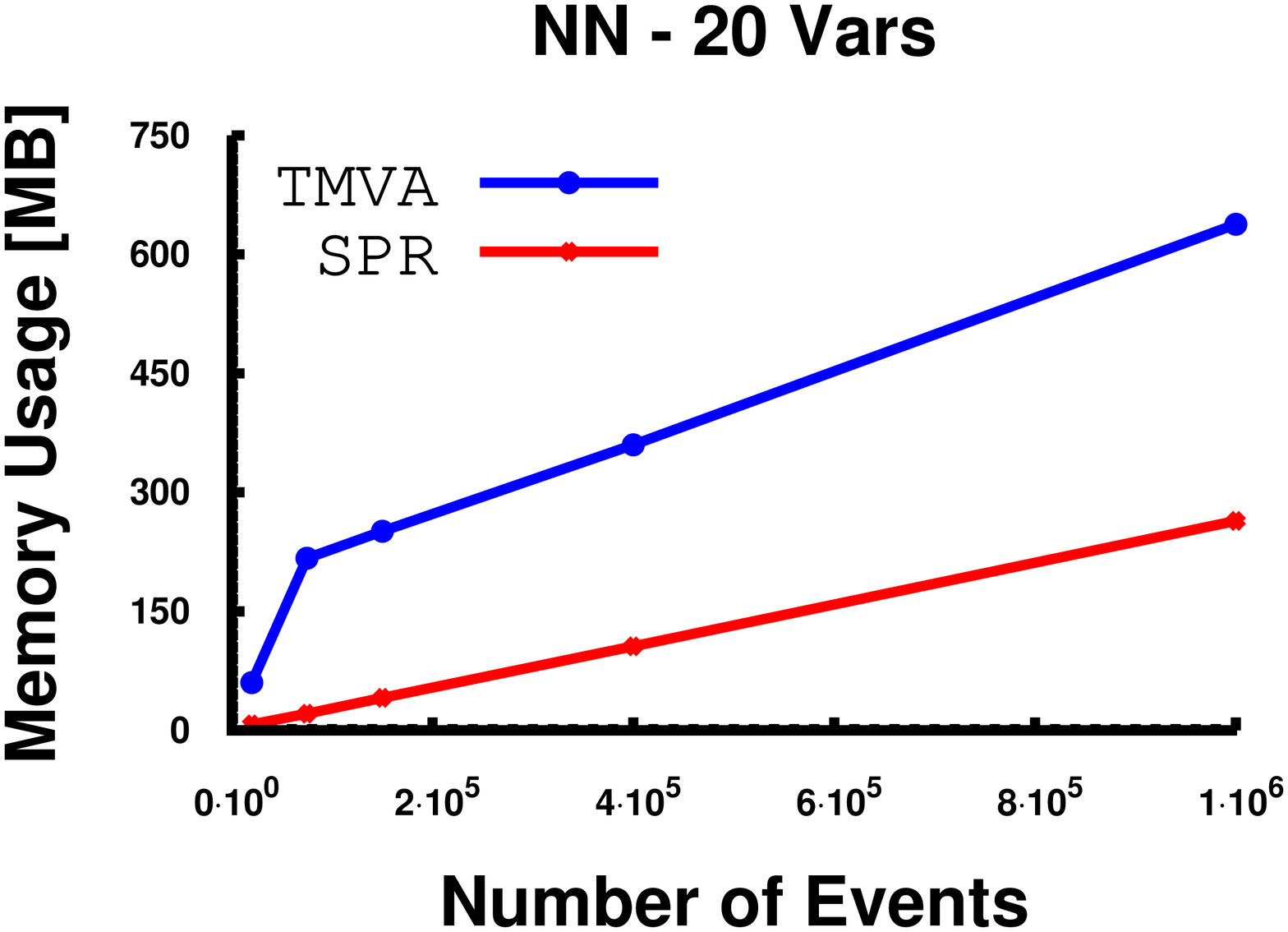}
\end{minipage}
\begin{minipage}[b] {0.5\linewidth} \centering
\includegraphics[scale=0.34, angle=270]{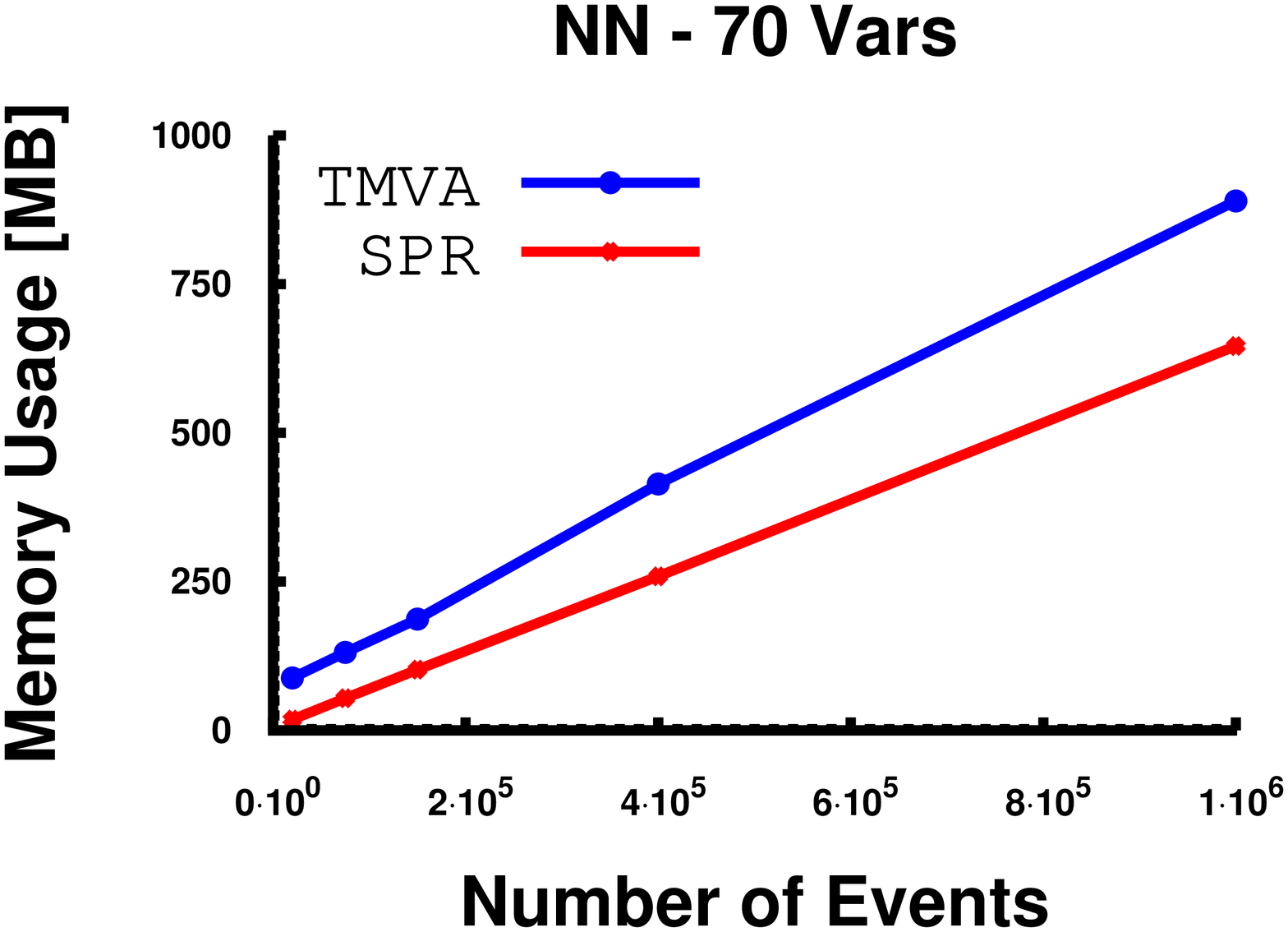}
\end{minipage}
\caption{Threenorm data set: NN memory usage vs number of events for 20 variables (left) and 70 variables (right).}
\label{fig:NormNNMemory}
\end{figure}

 Fig.~\ref{fig:NormNNMemory} shows the corresponding memory usage for the NN test. SPR consumes less memory for all the configurations  we tested.

\section{Test on ``Edge Case'' Data Sets}
\label{sec:4}

\begin{figure}[t]
\begin{minipage} [b]{0.5\linewidth}\centering
\includegraphics[scale=0.34, angle=270]{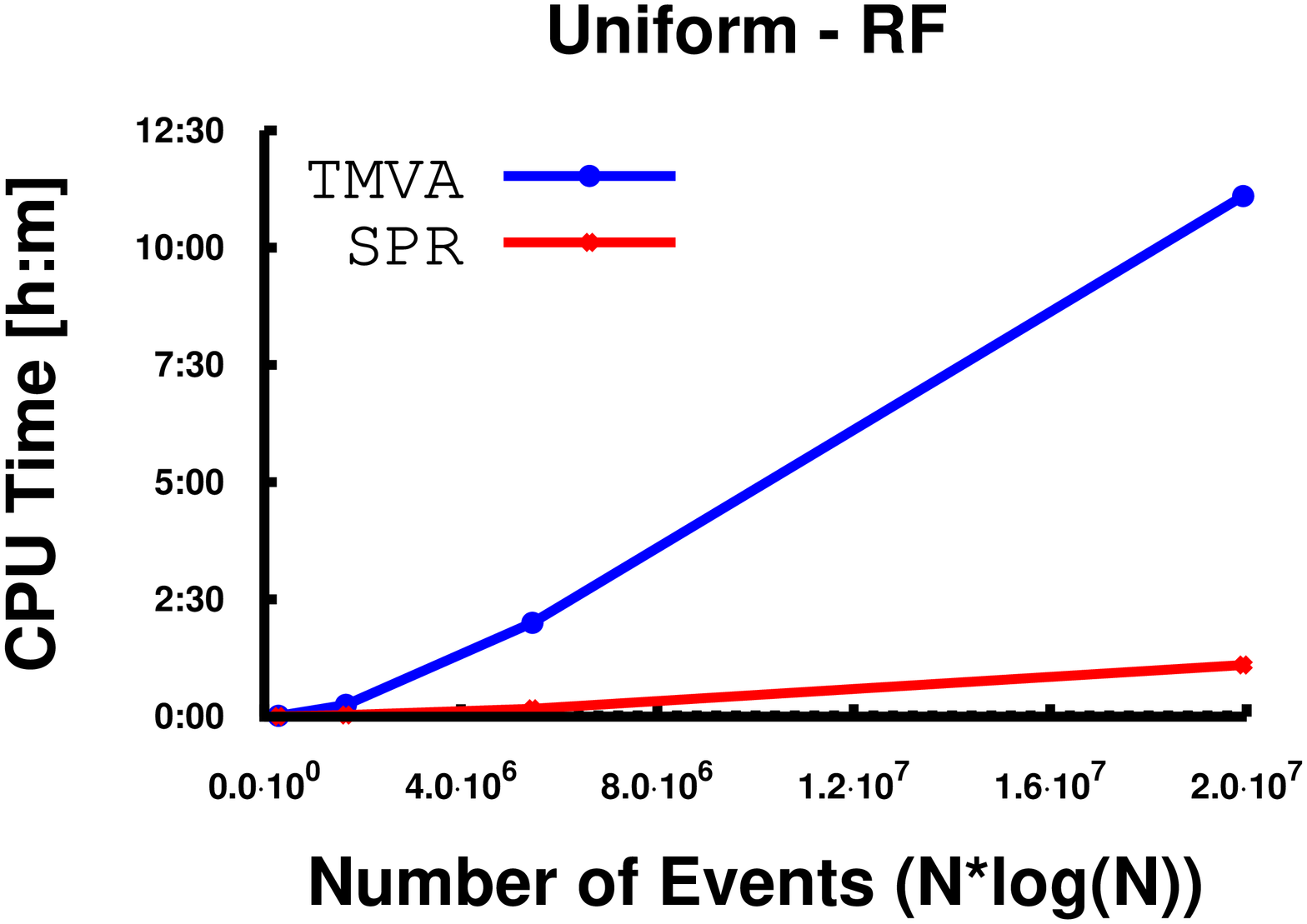}
\end{minipage}
\begin{minipage}[b] {0.5\linewidth} \centering
\includegraphics[scale=0.34, angle=270]{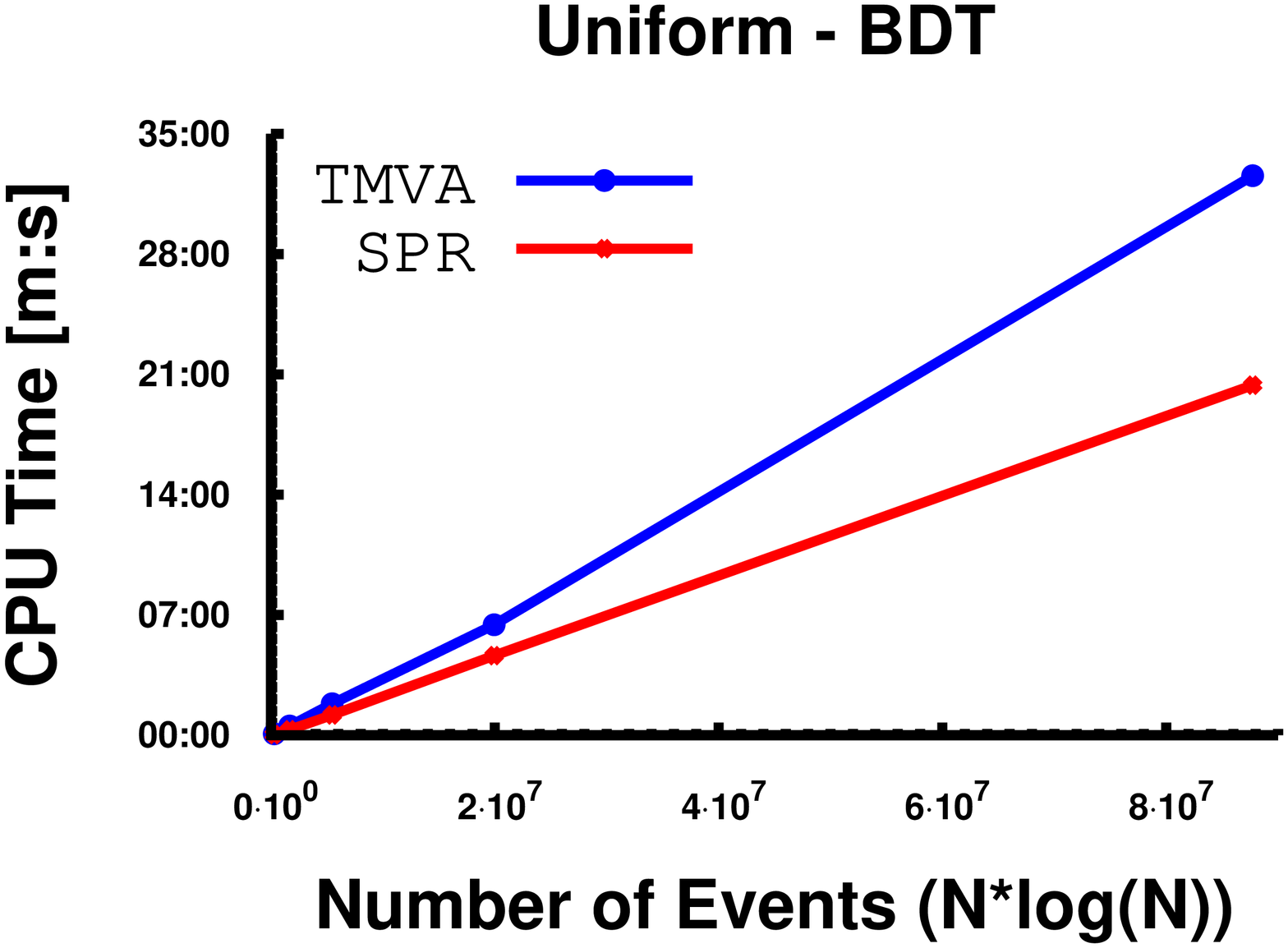}
\end{minipage}
\caption{``Uniform'' data set: RF (left) and BDT (right) learning CPU time vs number of events. }
\label{fig:UniformCPUBDT}
\end{figure}

We test SPR and TMVA on a data set with events characterized  by a single variable following the same uniform distribution for signal and background. We define this data set as ``Uniform''.
Fig.~\ref{fig:UniformCPUBDT} shows CPU time for RF (left) and BDT (right). The difference between SPR and TMVA for RF is extremely large. 
SPR is faster for BDT too. Again, RF builds large trees, while BDT trees are forced to be small by the optimization parameters.

\begin{figure}[t]
\begin{minipage} [b]{0.5\linewidth}\centering
\includegraphics[scale=0.34, angle=270]{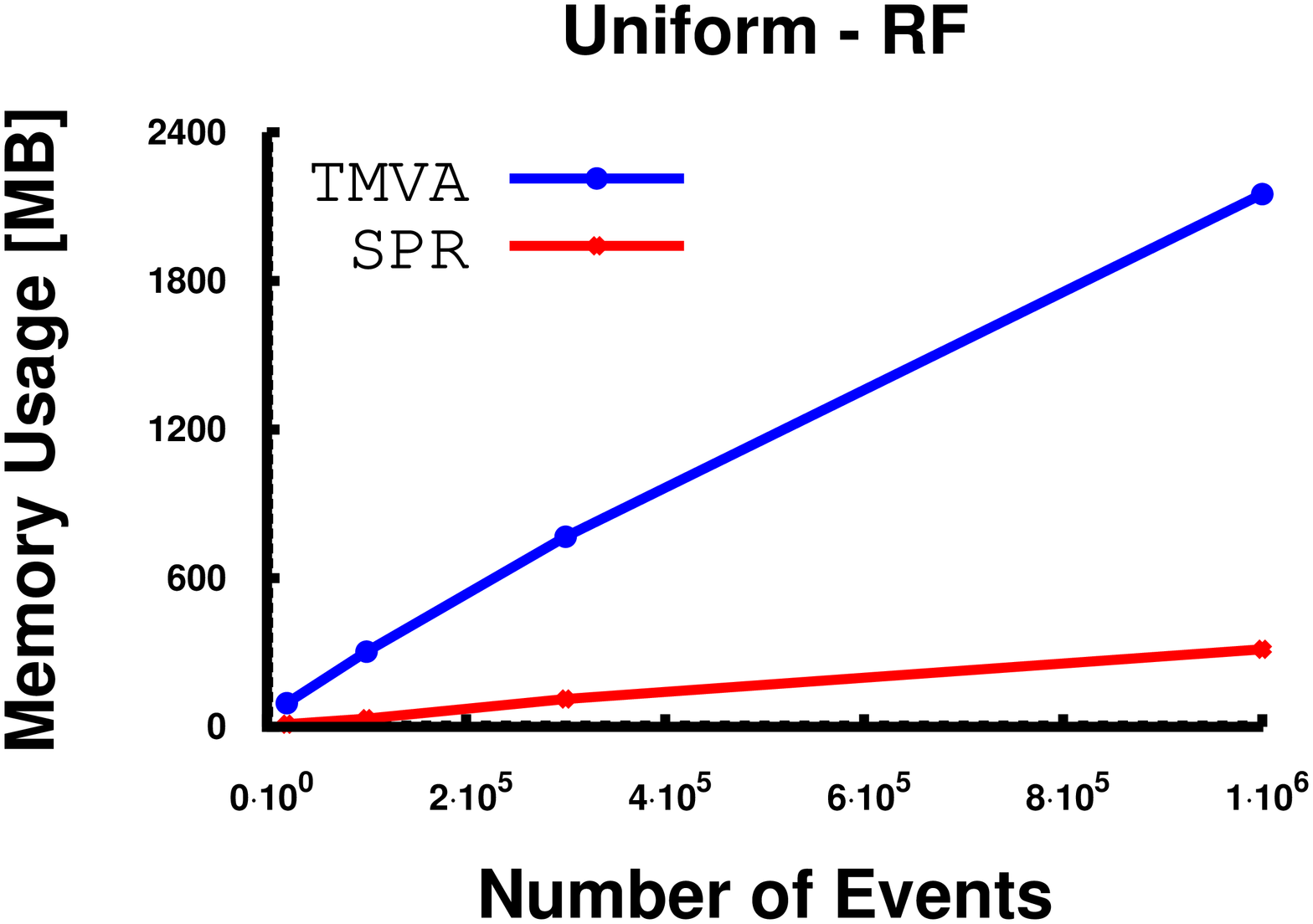}
\end{minipage}
\begin{minipage}[b] {0.5\linewidth} \centering
\includegraphics[scale=0.34, angle=270]{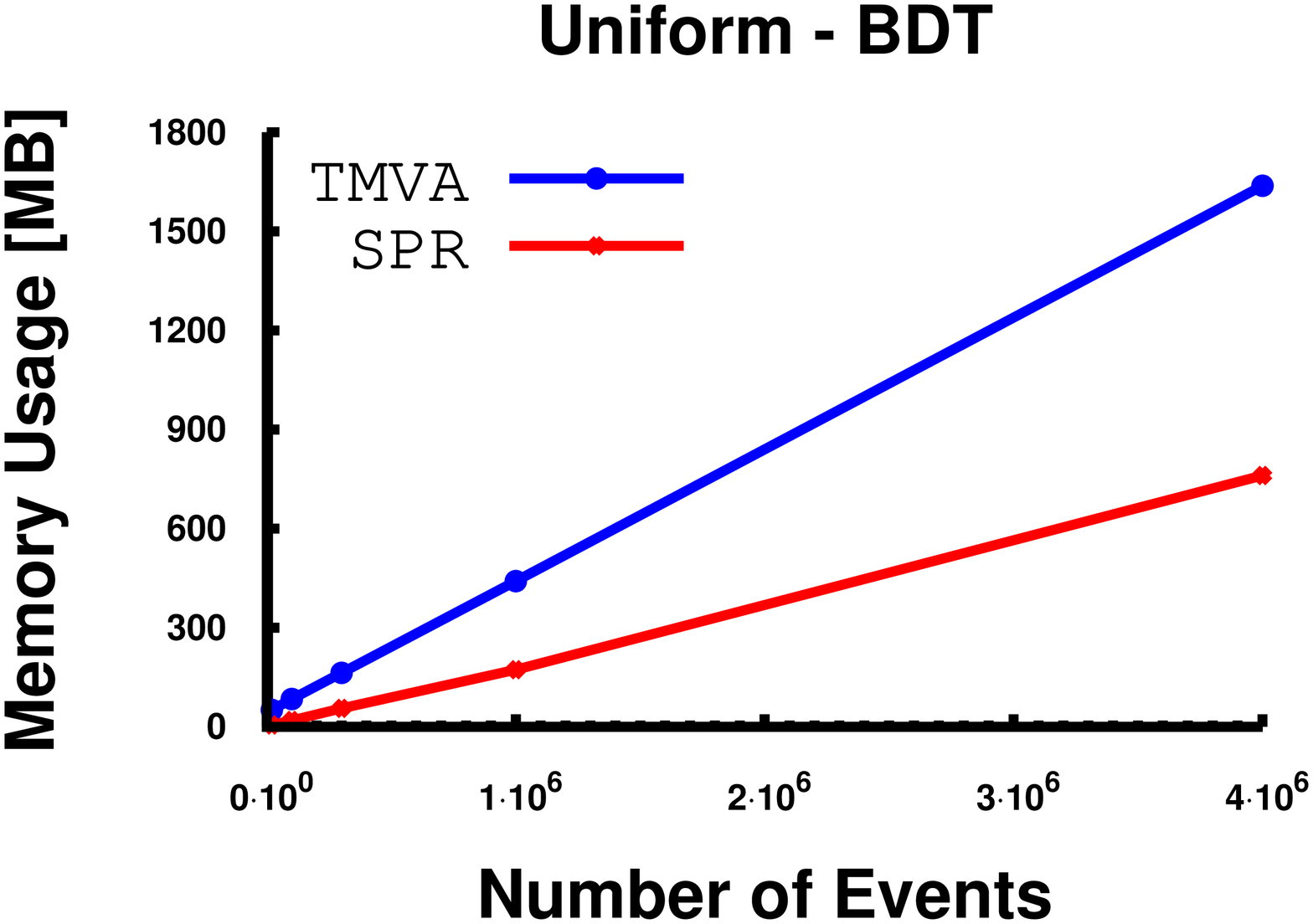}
\end{minipage}
\caption{``Uniform'' data set: RF (left) and BDT (right) memory usage vs number of events.}
\label{fig:Uniformemory}
\end{figure}

Fig.~\ref{fig:Uniformemory} shows TMVA and SPR memory usage for BDT and RF. Memory consumption mirrors CPU time. TMVA always consumes more memory than SPR.

\begin{figure}[t]
\begin{minipage} [b]{0.5\linewidth}\centering
\includegraphics[scale=0.34, angle=270]{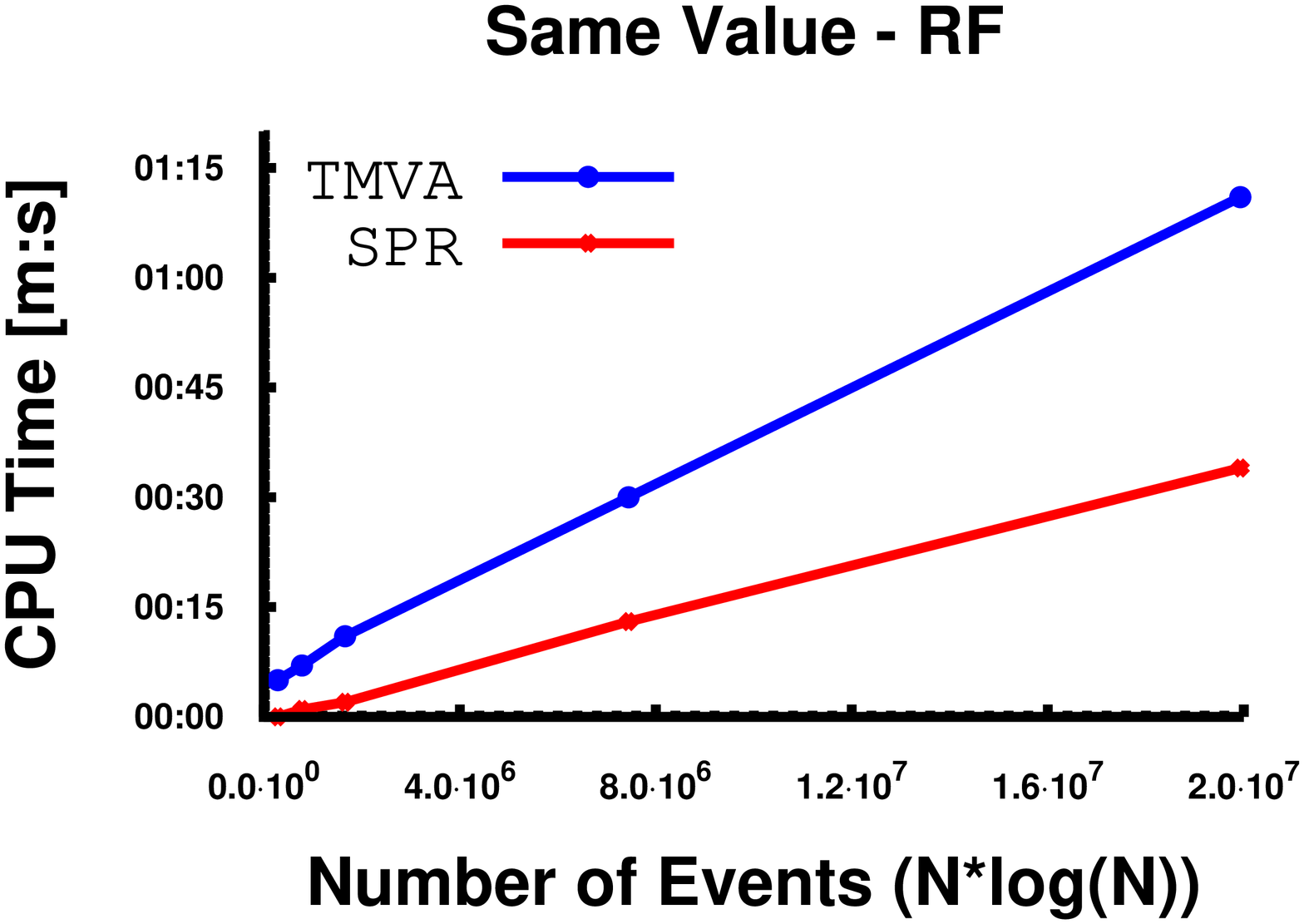}
\end{minipage}
\begin{minipage}[b] {0.5\linewidth} \centering
\includegraphics[scale=0.34, angle=270]{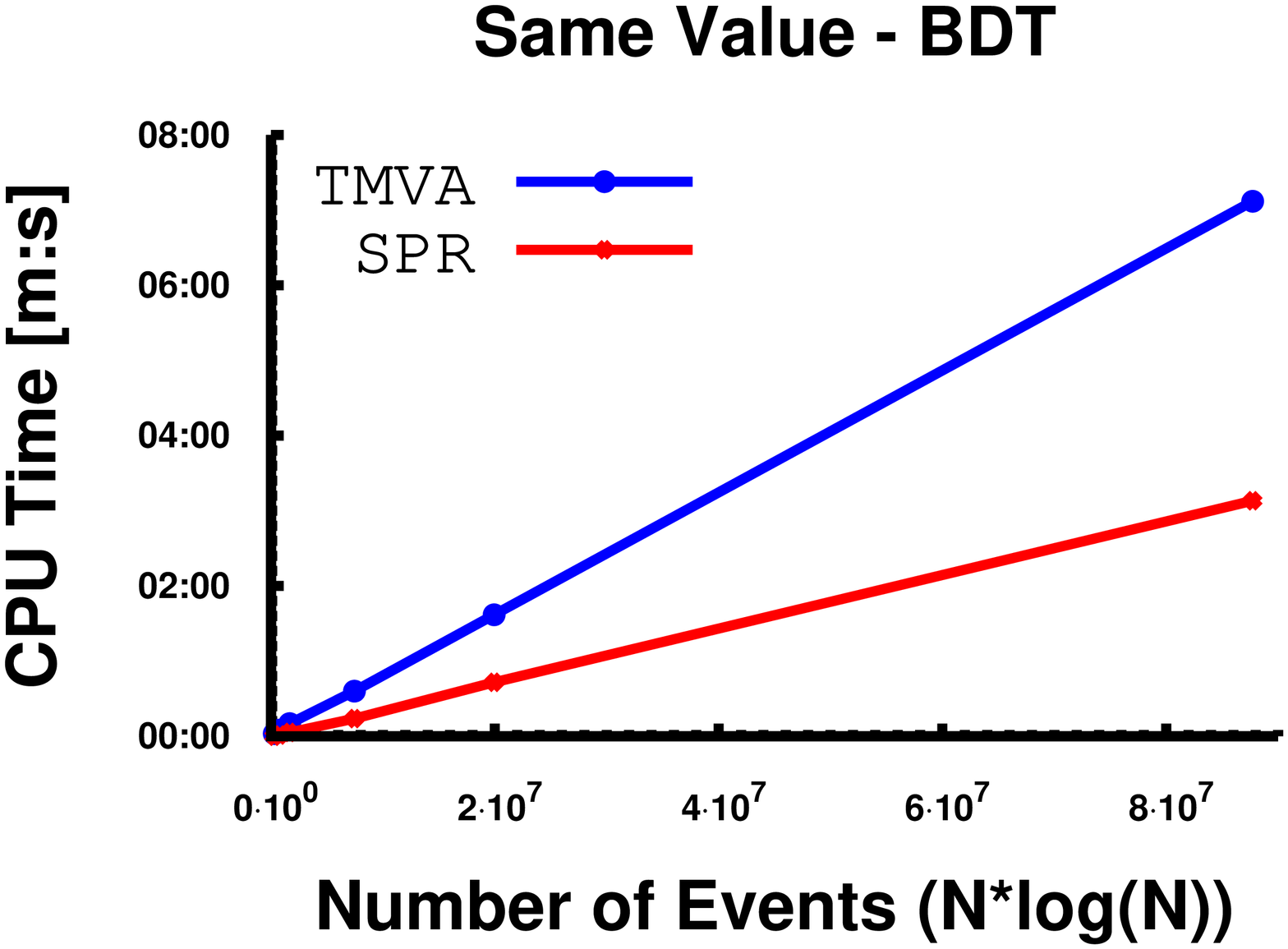}
\end{minipage}
\caption{``Same value'' data set: RF (left) and BDT (right) learning CPU time vs number of events.}
\label{fig:SameButOneCPU}
\end{figure}

Another edge case data set is characterized by a single variable that has the same value for all the events, except for one. We define it as ``Same value'' data set. As expected, learning time in this case is fast for both packages. CPU time outputs are shown in Fig.~\ref{fig:SameButOneCPU}. SPR is always faster than TMVA. In this test, SPR and TMVA grow similar trees for both RF and BDT.

\begin{figure}[t]
\begin{minipage} [b]{0.5\linewidth}\centering
\includegraphics[scale=0.34, angle=270]{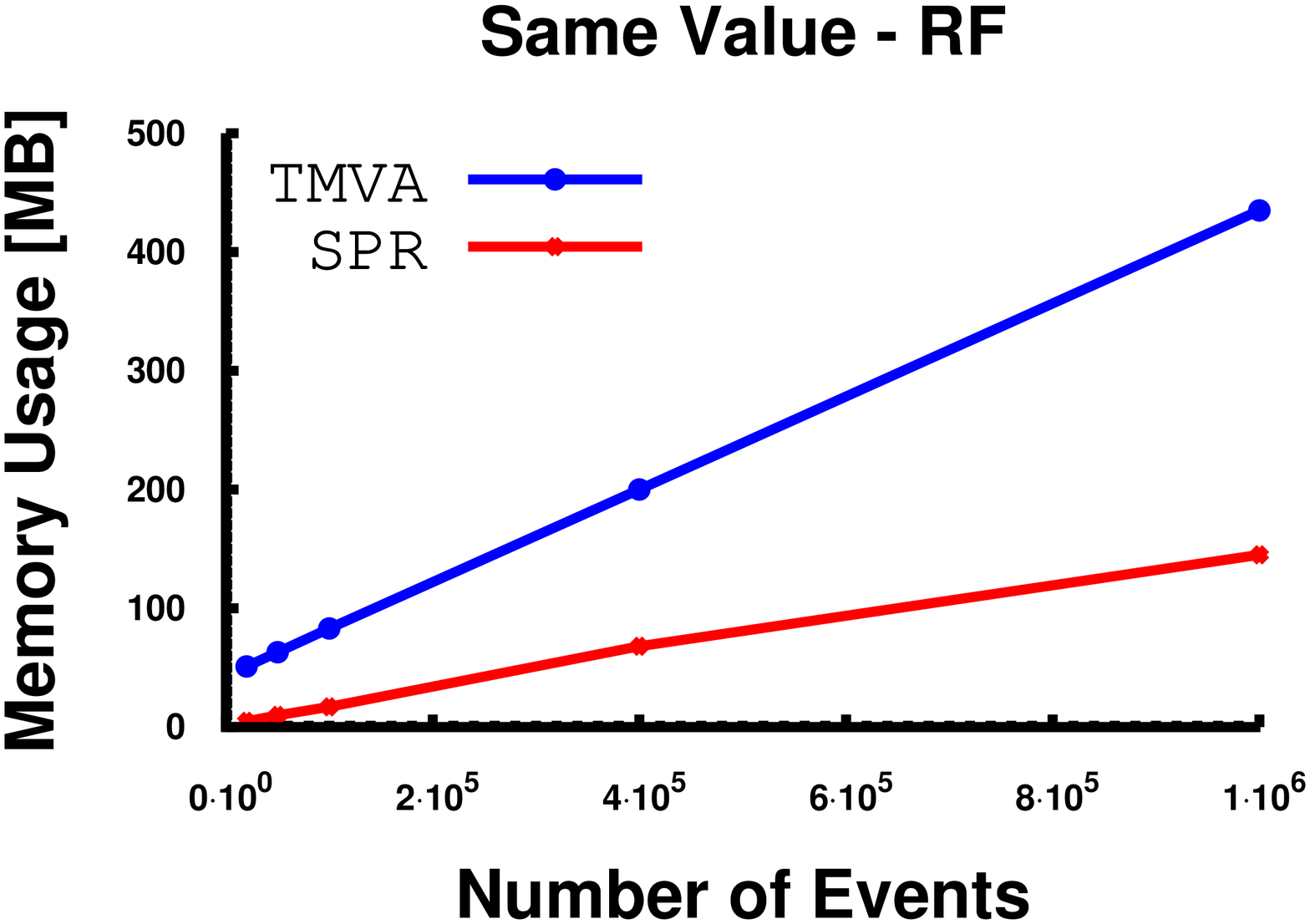}
\end{minipage}
\begin{minipage}[b] {0.5\linewidth} \centering
\includegraphics[scale=0.34, angle=270]{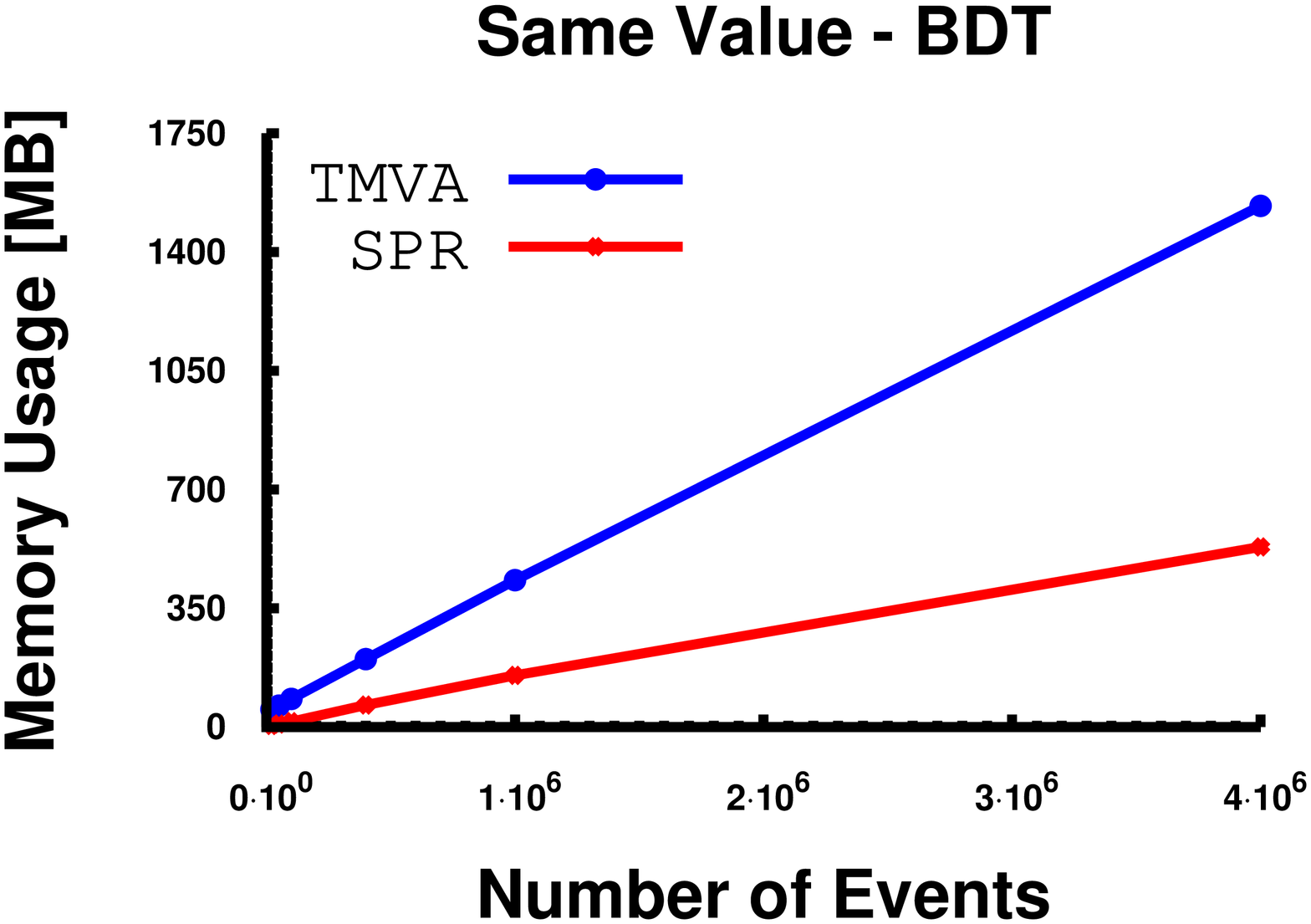}
\end{minipage}
\caption{``Same Value'' data set: RF (left) and BDT (right) memory usage vs number of events.}
\label{fig:SameButOneMemory}
\end{figure}

Fig.~\ref{fig:SameButOneMemory} shows TMVA and SPR memory usage for BDT and RF. TMVA  consumes constantly more memory than SPR for any classifier and any data set size.

\begin{figure}[t]
\begin{minipage} [b]{0.5\linewidth}\centering
\includegraphics[scale=0.34, angle=270]{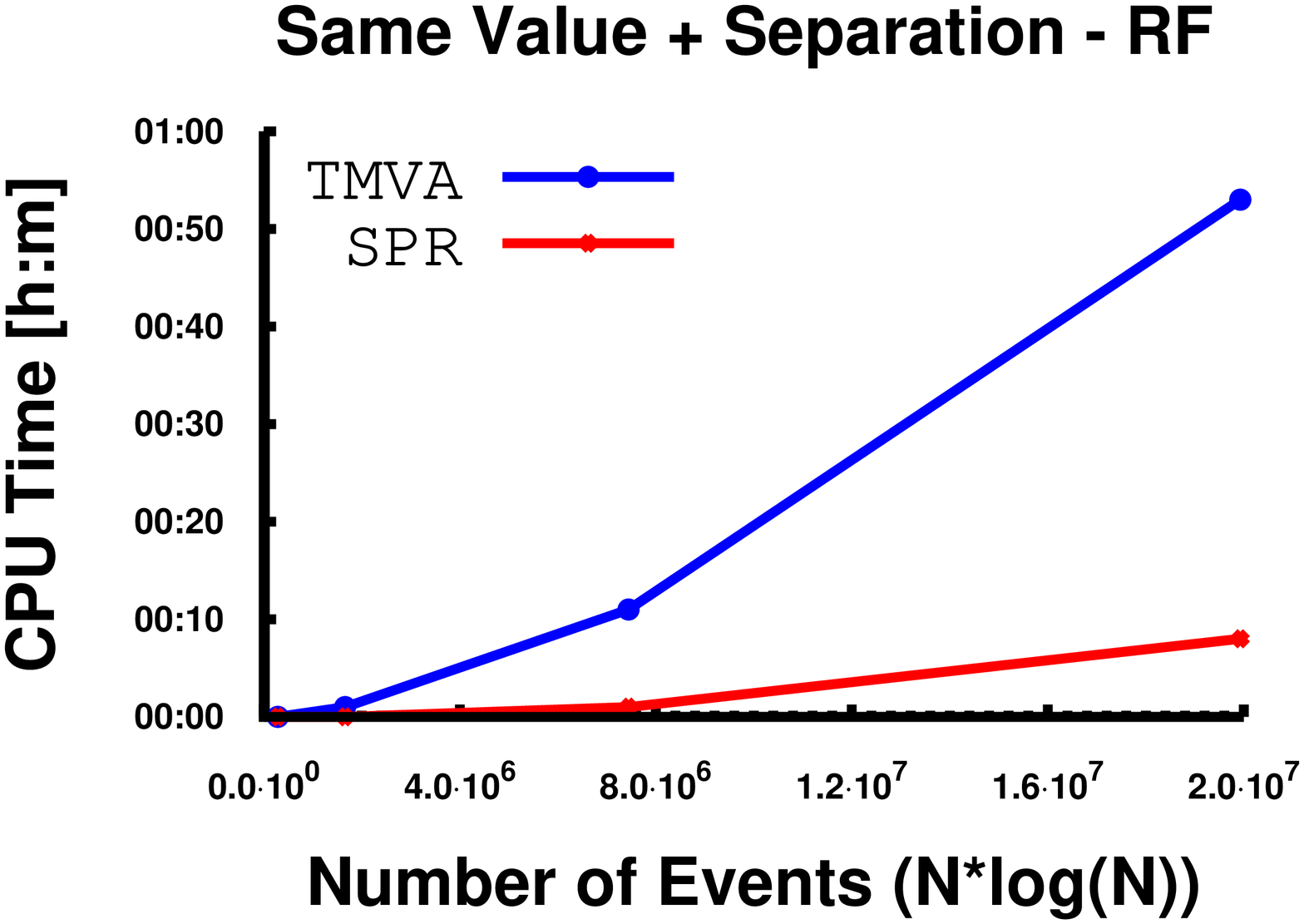}
\end{minipage}
\begin{minipage}[b] {0.5\linewidth} \centering
\includegraphics[scale=0.34, angle=270]{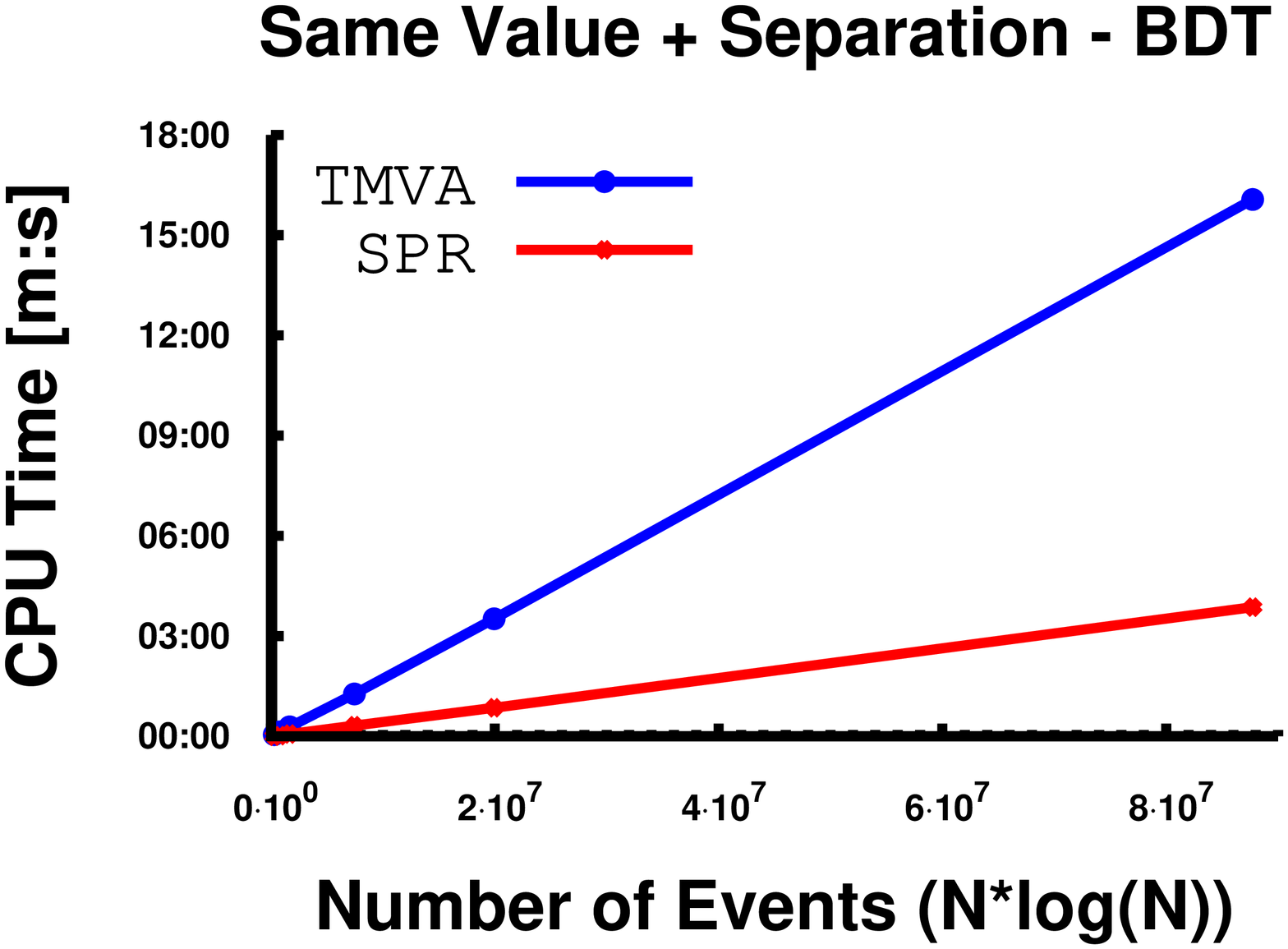}
\end{minipage}
\caption{``Same value and separation'' data set: RF (left) and BDT (right) learning CPU time vs number of events.}
\label{fig:SameSepCPU}
\end{figure}

\begin{figure}[t]
\begin{minipage} [b]{0.5\linewidth}\centering
\includegraphics[scale=0.34, angle=270]{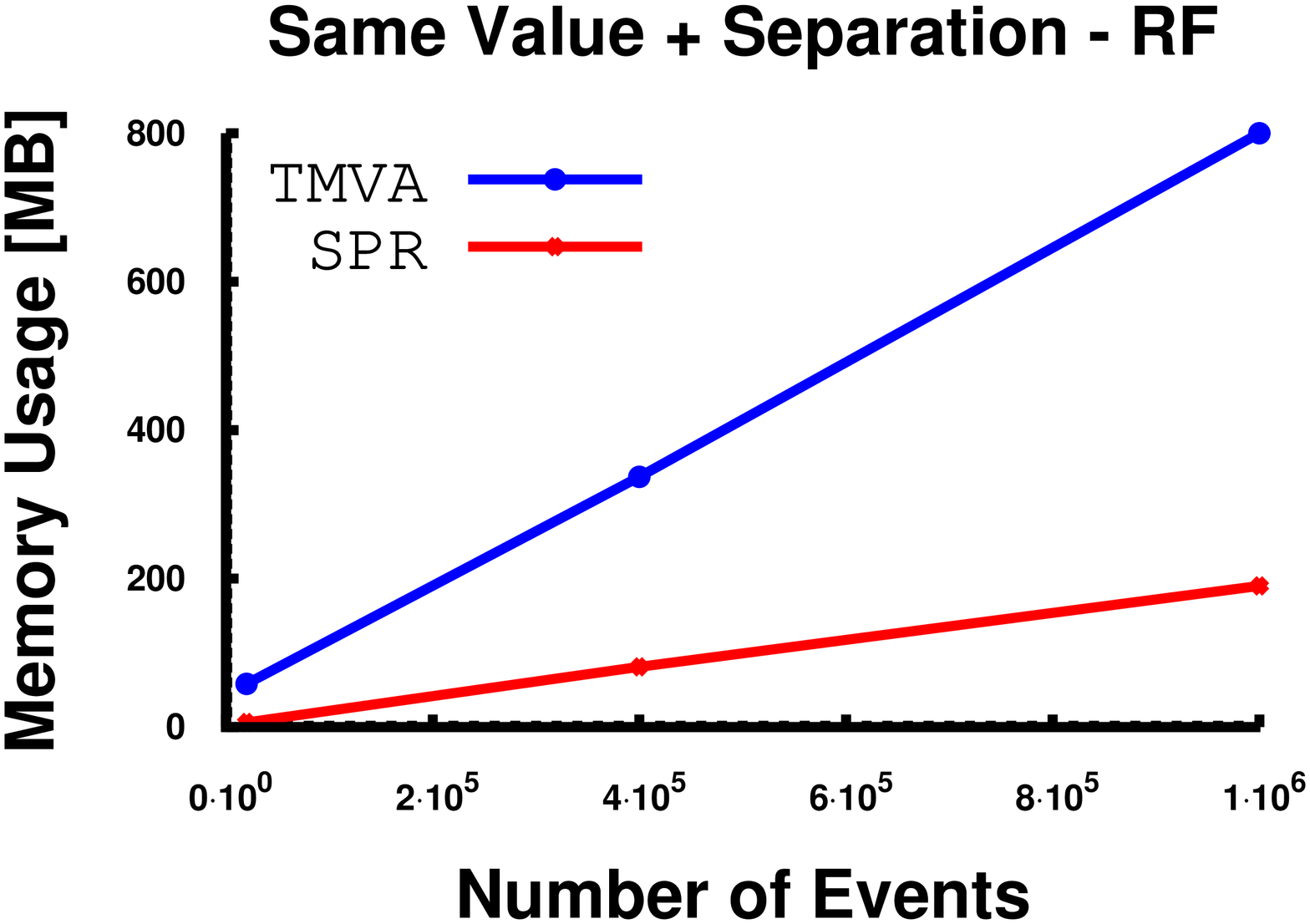}
\end{minipage}
\begin{minipage}[b] {0.5\linewidth} \centering
\includegraphics[scale=0.34, angle=270]{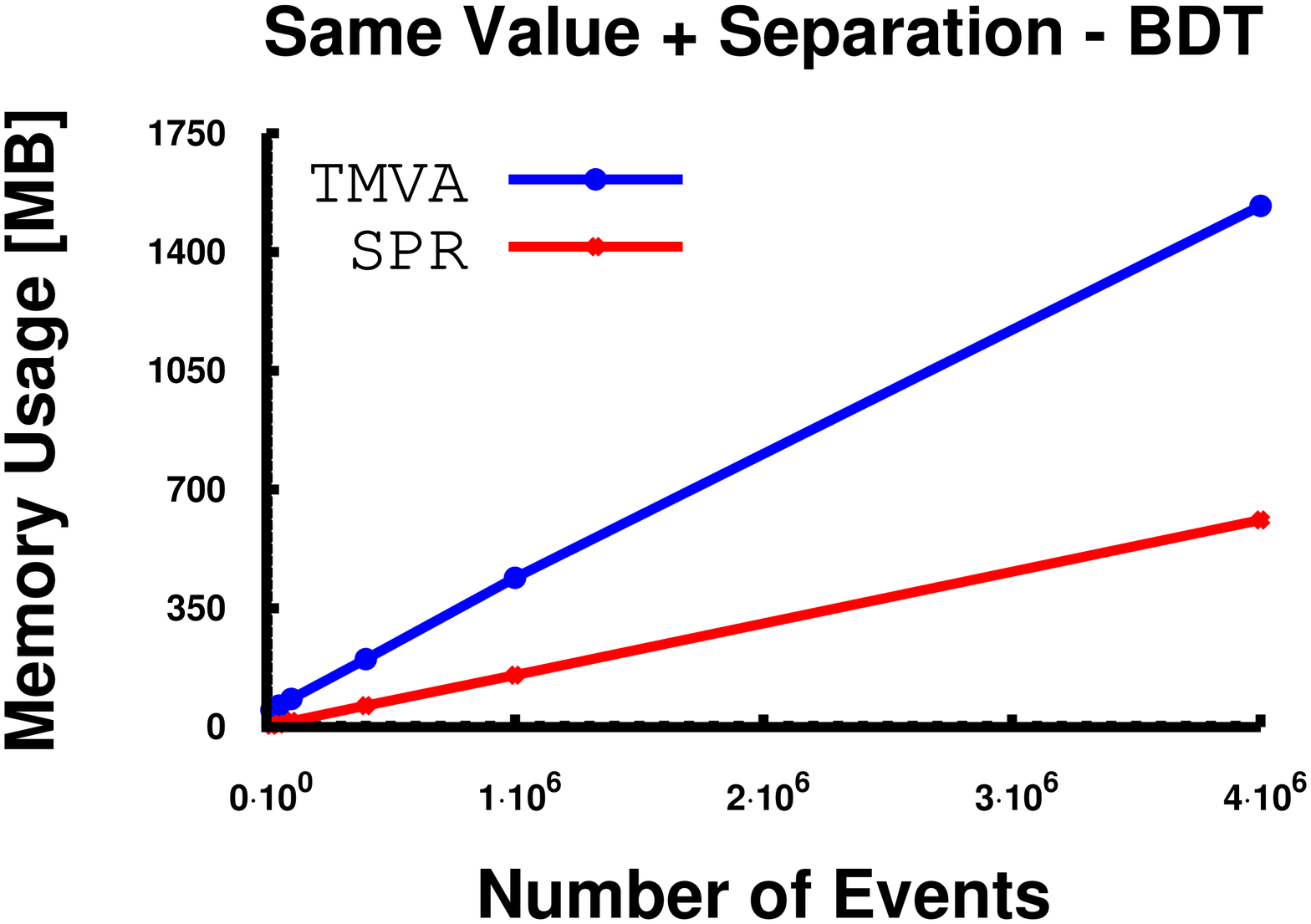}
\end{minipage}
\caption{``Same value and separation'' data set: RF (left) and BDT (right) memory usage vs number of events.}
\label{fig:SameSepMemory}
\end{figure}

SPR and TMVA are also tested on the ``Same value and separation'' data set, which is described by one variable that has the same value for 50\% of the  events and, for the remaining 50\%, follows the ``Threenorm'' distribution. That is, only half of the events are relevant to separate signal from background, with the remaining half being useless.  Results shown in Fig.~\ref{fig:SameSepCPU}  and Fig.~\ref{fig:SameSepMemory}   are in substantial agreement with the results  in  Sec. \ref{sec:3} and  Sec. \ref{sec:4}, with SPR always having a better performance than TMVA.

\section{Summary}
\label{sec:7}

We have evaluated CPU time and memory usage for different classifiers implemented in SPR and TMVA. As benchmark data, we used the Threenorm data set and targeted edge cases. In every single test we performed, SPR has been faster than TMVA and consumed less memory. For RF and BDT, the difference between SPR and TMVA  is particularly large.  NN performance of the two statistical packages is closer. For tree-based classifiers, the ratio between TMVA and SPR CPU time increased as the number of events increased.  In all other tests, the ratio between SPR and TMVA tended to be constant when increasing the data set size.

\section*{Acknowledgements}
Thanks to Ilya Narsky and Julian Bunn for insightful suggestions and reviewing earlier drafts of this paper.


\end{document}